\documentclass[twocolumn,showpacs,preprintnumbers,amsmath,amssymb]{revtex4}

\begin{document}
\title{The relationship between two flavors of oblivious transfer at the quantum
level}
\author{Guang-Ping He}
\affiliation{Department of Physics and Advanced Research Center,
Zhongshan University, Guangzhou 510275, China}
\author{Z. D. Wang}
\affiliation{Department of Physics, The University of Hong Kong,
Pokfulam Road, Hong Kong, China}

\begin{abstract}
Though all-or-nothing oblivious transfer and one-out-of-two
oblivious transfer are equivalent in classical cryptography, we
here show that due to the nature of quantum cryptography, a
protocol built upon secure quantum all-or-nothing oblivious
transfer cannot satisfy the rigorous definition of quantum
one-out-of-two oblivious transfer.
\end{abstract}

\pacs{03.67.Dd, 03.67.Hk, 89.70.+c} \maketitle

\newpage

\section{Introduction}

Mystery of quantum cryptography has long intrigued scientists.
 On one hand, several cryptographic tasks
such as the quantum conjugate coding\cite {Wiesner} and the
well-known quantum key distribution\cite{BB84,Ekert91,B92} have
made great successes. They achieved theoretically unbreakable
security which can never be reached by their classical
counterparts. But, on the other hand, some no-go theorems were
established, indicating that quantum cryptography is not always
powerful for any task. In particular, the MLC no-go theorem
\cite{Mayers,LF} rules out the possibility of non-relativistic
unconditionally secure quantum bit commitment (QBC), and the Lo's
insecurity proof of one-sided two-party quantum secure
computations\cite{impossible1} indicates that one-out-of-two
oblivious transfer is impossible either.

Oblivious transfer (OT) is an important concept found to be very
useful in designing multi-party cryptography
protocols\cite{Kilian}. There are two major flavours of OTs. The
original one\cite{Ra81,Wiesner} is simply known as oblivious
transfer, while sometimes can also be called all-or-nothing OT.
Another related notion was proposed later, which is called
one-out-of-two OT\cite{1-2OT}. In classical cryptography, it was
shown that these two are computationally equivalent\cite{p-OT}.
Essentially,  a protocol was presented in Ref.\cite{p-OT} to
illustrate that secure all-or-nothing OT can lead to secure
one-out-of-two OT. Furthermore, it was believed that secure
one-out-of-two OT can lead to secure BC\cite {impossible1}. This
standard classical reduction chain reveals the connection between
the security of OT and BC protocols in the classical level.

Very recently, a quantum all-or-nothing OT protocol was developed
\cite{QOT}. This OT does not rigorously satisfy the requirement of
one-sided two-party quantum secure computation protocols, on which
the Lo's insecurity proof was based. Thus it could remain
unconditionally secure against the cheating strategy in the Lo's
proof. Nevertheless, at the first glance, this result would
conflict with the Lo's conclusion and in turn with the MLC no-go
theorem (i.e., secure quantum one-out-of-two OT and QBC would be
possible) if the mentioned standard classical reduction were
justified.

More intriguingly, it has also been realized that ``reductions and
relations between classical cryptographic tasks need not
necessarily apply to their quantum equivalents''\cite{string BC}.
Indeed,  it will be shown in this paper that once we intend  to
build an one-out-of-two OT protocol  on a secure quantum
all-or-nothing OT protocol with the method developed in
Ref.\cite{p-OT}, it is impossible that the resultant protocol can
satisfy the rigorous definition of one-out-of-two OT on which the
Lo's proof was based. In this sense, secure quantum all-or-nothing
OT does not imply secure quantum one-out-of-two OT, i.e. the above
classical reduction chain is broken in the present quantum
cryptography case. As a result, there exists no logic conflict
between the existence of secure quantum all-or-nothing OT protocol
and 
the MLC no-go theorem of QBC.

The paper is organized as follows. In the next two sections, the
definitions of two flavors of OTs will be stated precisely and
 a brief review on their classical equivalence will be presented.
 The
 relationship between these OTs in the quantum level will be revealed in the
section IV, and how it is related to the cheating strategy in the
Lo's proof will be studied in the section V. In the section VI, it
will be indicated that the breaking of the reduction chain is not
simply a matter of the definition, rather it is originated from
the nature of quantum cryptography itself.

\section{Definitions}

Let us first state precisely the definitions of different OTs on
which the discussion in this paper is based. In Ref.\cite{p-OT}
where the classical equivalence between these OTs was proven, the
definitions of all-or-nothing OT and one-out-of-two OT were
summarized as:

\bigskip

\textit{Definition A: all-or-nothing OT}

(A-i) Alice knows one bit $b$.

(A-ii) Bob gets bit $b$ from Alice with probability $1/2$.

(A-iii) Bob knows whether he got $b$ or not.

(A-iv) Alice does not know whether Bob got $b$ or not.

\bigskip

\textit{Definition B: one-out-of-two OT}

(B-i) Alice knows two bits $b_{0}$ and $b_{1}$.

(B-ii) Bob gets bit $b_{j}$ and not $b_{\bar{j}}$ with $Pr(j=0)=Pr(j=1)=1/2$.

(B-iii) Bob knows which of $b_{0}$ or $b_{1}$ he got.

(B-iv) Alice does not know which $b_{j}$ Bob got.

\bigskip

In the Lo's insecurity proof of one-sided two-party quantum secure
computations\cite{impossible1}, a more rigorous definition of
one-out-of-two OT was specifically introduced as:

\bigskip

\textit{Definition C: rigorous one-out-of-two OT}

(C-i) Alice inputs $i$, which is a pair of messages $(m_{0},m_{1})$.

(C-ii) Bob inputs $j=0$ or $1$.

(C-iii) At the end of the protocol, Bob learns about the message $m_{j}$,
but not the other message $m_{\bar{j}}$, i.e., the protocol is an one-sided
two-party secure computation $f(m_{0},m_{1},j=0)=m_{0}$\ and $%
f(m_{0},m_{1},j=1)=m_{1}$.

(C-iv) Alice does not know which $m_{j}$ Bob got.

\bigskip

Meanwhile,  the definition of one-sided two-party quantum secure
computations used in the Lo's proof reads

\bigskip

\textit{Definition D: one-sided two-party secure computation}

Suppose Alice has a private (i.e. secret) input $i\in \{1,2,...,n\}$ and Bob
has a private input $j\in \{1,2,...,m\}$. Alice helps Bob to compute a
prescribed function $f(i,j)\in \{1,2,...,p\}$ in such a way that, at the end
of the protocol:

(a) Bob learns $f(i,j)$ unambiguously;

(b) Alice learns nothing [about $j$\ or $f(i,j)$];

(c) Bob knows nothing about $i$ more than what logically follows from the
values of $j$ and $f(i,j)$.

\bigskip

Obviously, Definition C is a special case of Definition D. In
Ref.\cite {impossible1} it is proven that any protocol satisfying
Definition D is insecure. Therefore as a corollary, there should
not exist a secure quantum one-out-of-two OT protocol which
satisfies Definition C rigorously.

\section{Classical equivalence}

The proof of the classical equivalence between the two flavors of
OTs is provided in Ref.\cite{1-2OT}. The major part of the proof
is the following procedure, showing how secure one-out-of-two OT
can be implemented upon secure all-or-nothing OT.

\bigskip

\textit{Protocol P:}

(1) Alice and Bob agree on a security parameter $s$;

(2) Alice chooses at random $Ks$ bits $r_{1},r_{2},...,r_{Ks}$;

(3) For each of these $Ks$ bits Alice uses the all-or-nothing OT protocol to
disclose the bit $r_{k}$ to Bob;

(4) Bob selects $U=\{i_{1},i_{2},...,i_{\alpha _{s}}\}$ and $V=\{i_{\alpha
_{s}+1},i_{\alpha _{s}+2},...,i_{2\alpha _{s}}\}$ where $\alpha _{s}=Ks/3$
with $U\cap V=\emptyset $ and such that he knows $r_{k_{l}}$ for each $%
k_{l}\in U$;

(5) Bob sends $(X,Y)=(U,V)$ or $(X,Y)=(V,U)$ to Alice according to a random
bit $j$;

(6) Alice computes $c_{0}=\bigoplus\limits_{x\in X}r_{x}$ and $%
c_{1}=\bigoplus\limits_{y\in Y}r_{y}$;

(7) Alice returns to Bob $b_{0}\oplus c_{0}$ and $b_{1}\oplus c_{1}$;

(8) Bob computes $\bigoplus\limits_{u\in U}r_{u}\in \{c_{0},c_{1}\}$ and
uses it to get his secret bit $b_{j}$.

\section{Relationship at the quantum level}

Though the two definitions of one-out-of-two OT (Definitions B and
C) seem to be consistent with each other, we here will show that,
in the quantum level, if a secure quantum all-or-nothing OT
protocol satisfies Definition A and can be used as a ``black
box'', a Protocol P built upon it via the above procedure does not
satisfy Definition C rigorously, though it satisfies Definition B.

The deviation from Definition C lies in (C-i) and (C-iii). Consider Alice's
input $i$ in Protocol P. In the step (7) of the protocol, we can see that $i$
includes not only the secret bits $b_{0}$ and $b_{1}$, but also $c_{0}$ and $%
c_{1}$. The steps (5) and (6) shows that $c_{0}$ and $c_{1}$ not only depend
on Alice's input $r_{1},r_{2},...,r_{Ks}$, but also depend on how Bob
selects $X$, $Y$, $U$ and $V$, i.e. they depend on Bob's input $j$.
Therefore, Protocol P cannot be viewed as a ``black box'' function $%
f(i(m_{0},m_{1}),j)$, where $i$ and $j$ are the private inputs of
Alice and Bob respectively. Instead, it has the form
$f(i(m_{0},m_{1},j),j)$, where Alice' input will be varied
according to Bob's input, and its value is not determined until
Bob's input has been completed. That is, Protocol P does not
rigorously satisfy Definition C, nor Definition D as the
description of the function $f$ is different.

Though the difference seems tiny at the first glance, its consequences are
significant at the quantum level. This can be seen from two aspects:

\textit{(I) The con side: Protocol P cannot be used as a black
box} since the sequence of the participants' inputs is important,
i.e. we have to deal with the details of the protocol when  it is
used to build other protocols. As argued in the introduction of
Ref.\cite{impossible1}, to ensure that the  standard classical
reduction can apply to quantum cryptographic protocols, ``one must
be allowed to use a quantum cryptographic protocol as a `black
box' primitive in building up more sophisticated protocols and to
analyze the security of those new protocols with classical
probability theory''. Therefore the above character of Protocol P
make it unsuitable to be used as a rigorous quantum one-out-of-two
OT to connect the reduction chain between quantum all-or-nothing
OT and QBC. Other applications of Protocol P in quantum
cryptography may also have a limited power.

\textit{(II) The pro side: Protocol P is not covered by the cheating
strategy in Ref.\cite{impossible1}} for the following reason. According to
the strategy, Bob can change the value of $j$ from $j_{1}$ to $j_{2}$ by
applying a unitary transformation to his own quantum machine. Therefore he
can learn $f(i(m_{0},m_{1}),j_{1})$\ and $f(i(m_{0},m_{1}),j_{2})$\
simultaneously without being found by Alice. However, for the function $%
f(i(m_{0},m_{1},j),j)$, the value $f(i(m_{0},m_{1},j_{1}),j_{2})$\ is
meaningless. Without the help of Alice, Bob cannot change $i$ from $%
i(m_{0},m_{1},j_{1})$\ to $i(m_{0},m_{1},j_{2})$. Hence he cannot learn $%
f(i(m_{0},m_{1},j_{1}),j_{1})$\ and $f(i(m_{0},m_{1},j_{2}),j_{2})$\
simultaneously by\ himself. Namely, though the cheating strategy works for
any protocol satisfying Definition D, it does not work for Protocol P.

On the other hand, though $c_{0}$ and $c_{1}$ depend on Bob's
input $j$, from the protocol it can be seen clearly that they are
insufficient for Alice to learn the value of $j$. Thus Protocol P
is still secure against Alice. In this sense, the relaxed
definition of one-out-of-two OT (Definition B) is satisfied.

\section{Defeating the cheating strategy}

In this section, the above conclusion (II) will be rigorously
proven. For convenience, let us first recall the cheating strategy
in the Lo's proof in more details. According to the section III of
Ref.\cite{impossible1}, in any protocol satisfying Definition D,
Alice and Bob's actions on their quantum machines can be
summarized as an overall unitary transformation $U$ applied to the
initial state $\left| u\right\rangle _{in}\in H_{A}\otimes H_{B}$,
i.e.
\begin{equation}
\left| u\right\rangle _{fin}=U\left| u\right\rangle _{in}.  \label{e1}
\end{equation}
When both parties are honest, $\left| u^{h}\right\rangle _{in}=\left|
i\right\rangle _{A}\otimes \left| j\right\rangle _{B}$ and
\begin{equation}
\left| u^{h}\right\rangle _{fin}=\left| v_{ij}\right\rangle \equiv U(\left|
i\right\rangle _{A}\otimes \left| j\right\rangle _{B}).  \label{e2}
\end{equation}
Therefore the density matrix that Bob has at the end of protocol is
\begin{equation}
\rho ^{i,j}=Tr_{A}\left| v_{ij}\right\rangle \left\langle v_{ij}\right| .
\label{e3}
\end{equation}

Bob can cheat in this protocol, because given $j_{1},j_{2}\in \{1,2,...,m\}$%
, there exists a unitary transformation $U^{j_{1},j_{2}}$ such that
\begin{equation}
U^{j_{1},j_{2}}\rho ^{i,j_{1}}(U^{j_{1},j_{2}})^{-1}=\rho ^{i,j_{2}}
\label{e4}
\end{equation}
for all $i$. It means that Bob can change the value of $j$ from $j_{1}$ to $%
j_{2}$ by applying a unitary transformation independent of $i$\ to the state
of his quantum machine. This equation is proven as follows.

Alice may entangles the state of her quantum machine $A$ with her quantum
dice $D$ and prepares the initial state
\begin{equation}
\frac{1}{\sqrt{n}}\sum\limits_{i}\left| i\right\rangle _{D}\otimes \left|
i\right\rangle _{A}.  \label{e5}
\end{equation}
She keeps $D$ for herself and uses the second register $A$ to
execute the protocol. Suppose Bob's input is $j_{1}$. The initial
state is
\begin{equation}
\left| u^{\prime }\right\rangle _{in}=\frac{1}{\sqrt{n}}\sum\limits_{i}%
\left| i\right\rangle _{D}\otimes \left| i\right\rangle _{A}\otimes \left|
j_{1}\right\rangle _{B}.  \label{e6}
\end{equation}
At the end of the protocol, it follows from Eqs.(\ref{e1}) and (\ref{e6})
that the total wave function of the combined system $D$, $A$, and $B$ is
\begin{equation}
\left| v_{j_{1}}\right\rangle _{in}=\frac{1}{\sqrt{n}}\sum\limits_{i}\left|
i\right\rangle _{D}\otimes U(\left| i\right\rangle _{A}\otimes \left|
j_{1}\right\rangle _{B}).  \label{e7}
\end{equation}
Similarly, if Bob's input is $j_{2}$, the total wave function at the end
will be
\begin{equation}
\left| v_{j_{2}}\right\rangle _{in}=\frac{1}{\sqrt{n}}\sum\limits_{i}\left|
i\right\rangle _{D}\otimes U(\left| i\right\rangle _{A}\otimes \left|
j_{2}\right\rangle _{B}).  \label{e8}
\end{equation}
Due to the requirement (b) in Definition D, the reduced density matrices in
Alice's hand for the two cases $j=j_{1}$ and $j=j_{2}$ must be the same,
i.e.
\begin{equation}
\rho _{j_{1}}^{Alice}=Tr_{B}\left| v_{j_{1}}\right\rangle \left\langle
v_{j_{1}}\right| =Tr_{B}\left| v_{j_{2}}\right\rangle \left\langle
v_{j_{2}}\right| =\rho _{j_{2}}^{Alice}.  \label{e9}
\end{equation}
Equivalently, $\left| v_{j_{1}}\right\rangle $ and $\left|
v_{j_{2}}\right\rangle $\ have the same Schmidt decomposition
\begin{equation}
\left| v_{j_{1}}\right\rangle =\sum\limits_{k}a_{k}\left| \alpha
_{k}\right\rangle _{AD}\otimes \left| \beta _{k}\right\rangle _{B}
\label{e10}
\end{equation}
and
\begin{equation}
\left| v_{j_{2}}\right\rangle =\sum\limits_{k}a_{k}\left| \alpha
_{k}\right\rangle _{AD}\otimes \left| \beta _{k}^{\prime }\right\rangle _{B}.
\label{e11}
\end{equation}
Now consider the unitary transformation $U^{j_{1},j_{2}}$ that rotates $%
\left| \beta _{k}\right\rangle _{B}$\ to $\left| \beta _{k}^{\prime
}\right\rangle _{B}$. Notice that it acts on $H_{B}$ alone and yet, as can
be seen from Eqs.(\ref{e10}) and (\ref{e11}), it rotates $\left|
v_{j_{1}}\right\rangle $ to $\left| v_{j_{2}}\right\rangle $,\ i.e.
\begin{equation}
\left| v_{j_{2}}\right\rangle =U^{j_{1},j_{2}}\left| v_{j_{1}}\right\rangle .
\label{e12}
\end{equation}
Since
\begin{equation}
_{D}\left\langle i\right. \left| v_{j}\right\rangle =\frac{1}{\sqrt{n}}%
\left| v_{ij}\right\rangle  \label{e13}
\end{equation}
[see Eqs.(\ref{e2}), (\ref{e7}), and (\ref{e8})], by multiplying Eq.(\ref
{e12}) by $_{D}\left\langle i\right| $\ on the left, one finds that
\begin{equation}
\left| v_{ij_{2}}\right\rangle =U^{j_{1},j_{2}}\left|
v_{ij_{1}}\right\rangle .  \label{e14}
\end{equation}
Taking the trace of $\left| v_{ij_{2}}\right\rangle \left\langle
v_{ij_{2}}\right| $\ over $H_{A}$ and using Eq.(\ref{e14}),
Eq.(\ref{e4}) can be obtained.

Note that all these equations are just those presented in  the
Lo's proof \cite{impossible1}. We now consider Protocol P, where
Alice's input $i$\ is dependent of Bob's input $j$. In the above
proof, all $i$ in the equations should be replaced by $i(j)$ from
the very beginning. Consequently, Eq.(\ref{e13}) becomes
\begin{equation}
_{D}\left\langle i(j)\right| \left. v_{j}\right\rangle =\frac{1}{\sqrt{n}}%
\left| v_{i(j)j}\right\rangle .  \label{e15}
\end{equation}
In this case multiplying Eq.(\ref{e12}) by $_{D}\left\langle
i_{2}\right| $\ ($i_{2}\equiv i(j_{2})$ for short) on the left
cannot give Eq.(\ref{e14}) any more. Instead, the result is
\begin{equation}
\left| v_{i_{2}j_{2}}\right\rangle =U^{j_{1},j_{2}}U^{i_{1},i_{2}}\left|
v_{i_{1}j_{1}}\right\rangle ,  \label{e16}
\end{equation}
where $U^{i_{1},i_{2}}\equiv _{D}\left| i_{2}\right\rangle \left\langle
i_{1}\right| _{D}$. Then Eq.(\ref{e4}) is replaced by
\begin{equation}
U^{j_{1},j_{2}}U^{i_{1},i_{2}}\rho
^{i_{1},j_{1}}(U^{j_{1},j_{2}}U^{i_{1},i_{2}})^{-1}=\rho ^{i_{2},j_{2}}.
\label{e17}
\end{equation}
Note that $U^{i_{1},i_{2}}$ is the unitary operation on Alice's side. This
implies that without Alice's help, Bob cannot change the density matrix he
has from $\rho ^{i_{1},j_{1}}$\ to $\rho ^{i_{2},j_{2}}$. That is why Bob's
cheating strategy fails in Protocol P.

\section{Origin of the inequivalence}

It is valuable to find out the underlying reason why Protocol P
does not satisfy the rigorous Definition C.  An illusion is
naturally aroused that the reason is due to a relaxed Definition A
of all-or-nothing OT used in the work. However, it is not true. In
fact, we never need to deal with the details of the all-or-nothing
OT in the section IV; we simply use it as a black box. Even when
the most rigorous definition of all-or-nothing OT is used, the
discussion in that section is still valid. Thus it is not a matter
of definition that the classical equivalence between the two
flavours of OTs cannot rigorously apply to the present quantum
case.

The real origin of this result can be found in the equations in
the previous section. By comparing Eqs.(\ref{e13}) and
(\ref{e15}), we can see that if there does not exist a system $D$,
Protocol P will become insecure too. That is, if Alice does not
introduce the quantum system $D$ in Eq.(\ref{e5}), Protocol P will
show no difference from the protocols satisfying Definition D. In
classical cryptography, Alice surely does not have such a system.
That is why the two flavors of OTs seem equivalent. In quantum
cryptography, if Alice does not make full use of the computational
power but simply executes the protocol with the quantum system $A$
alone, she cannot defeat Bob's cheating either. The difference
between Protocol P and a rigorous one-out-of-two OT can only be
manifested when the protocol is indeed executed at the quantum
level. In this sense, the underlying origin is the nature of
quantum cryptography itself.

\section{Discussions and summary}

It has been shown that though one-out-of-two OT can be built upon
all-or-nothing OT in classical cryptography, a Protocol P built
upon a secure quantum all-or-nothing OT protocol via the same
method cannot satisfy the rigorous Definition C of quantum
one-out-of-two OT.
 Considering that a secure quantum
all-or-nothing OT protocol was already established \cite{QOT},
which is not denied by the Lo's insecurity proof of the one-sided
two-party secure computations\cite{impossible1} because it does
not satisfy the requirement on which the proof is based,
it seems unlikely that such a protocol can lead to another
protocol satisfying the requirement. Furthermore, if a secure
protocol satisfying the rigorous definition of quantum
one-out-of-two OT  existed,  it would be used as a black box
primitive to implement secure QBC according to
Ref.\cite{impossible1}, conflicting with
the MLC no-go theorem. On the contrary, it is more logically
consistent that  no other method is available to build a rigorous
quantum one-out-of-two OT protocol upon quantum all-or-nothing OT.
That is, the two flavors of OTs should not be rigorously
equivalent in quantum cryptography.

Though the profound understanding of the exact relationship
between the two flavors of OTs at the quantum level is still
awaited, at least, one thing is clearly elaborated in this work:
 the classical equivalence between these OTs
cannot be directly applied to quantum cryptography. This finding
provides yet an intriguing example demonstrating that reductions
and relations between classical cryptographic tasks need careful
re-examination in  quantum cases.

\end{document}